\documentclass[11pt,letterpaper]{article}
\usepackage[margin=1in]{geometry}
\usepackage{ifpdf}
\ifpdf
    \usepackage[pdftex]{graphicx}
    \graphicspath{{figs/}{matlab/}}   
    \usepackage[update]{epstopdf}
\else
	\usepackage{graphicx}
	\graphicspath{{figs/}{matlab/}}   
\fi
\usepackage{pdflscape}
\usepackage{wrapfig,bbm}
\usepackage{enumitem,color}
\usepackage{amsthm}
\newtheorem{theorem}{Theorem}

\usepackage{hyperref}
\usepackage{array}
\usepackage{amsmath,amsthm}
\usepackage{amssymb}
\usepackage{amstext}
\usepackage{cite,setspace}
\usepackage{multirow}

\begin{document}

\title{A Note on the Fundamental Limits of Coded Caching}
\author{Chao Tian}
\maketitle

\begin{abstract}
The fundamental limit of coded caching is investigated for the case with $N=3$ files and $K=3$ users. An improved outer bound is obtained through the computational approach developed by the author in an earlier work. This result is part of the online collection of ``Solutions of Computed Information Theoretic Limits (SCITL)''. 
\end{abstract}

\section{Introduction}

In a recent work \cite{MaddahAliNiesen:14}, Maddah-Ali and Niesen considered the caching problem which deals with improving the content delivery efficiency, {\em i.e.}, the total traffic rate $R$ in the delivery phase, through the utilization of local cache memory of capacity $M$ each where the coded contents are strategically prepared in the placement phase, in a system with $N$ files and $K$ users. It was shown coded caching can be rather beneficial, and in fact orderly optimal, while uncoded caching solution suffers a significant loss. Subsequent works extended it to decentralized caching placements \cite{MaddahAliNiesen:14Networking}, caching with nonuniform demands \cite{MaddahAliNiesen:14infocom}, and online caching placements \cite{Pedarsani:14}, etc.. 

Despite these advances, the fundamental tradeoff between the delivery traffic rate $R$ and cache memory capacity $M$ is not fully known except the two-user two-file case. This is partly due to the fact that the main focus of these existing investigations \cite{MaddahAliNiesen:14,MaddahAliNiesen:14Networking,MaddahAliNiesen:14infocom,Pedarsani:14} is on the regime when the number of files and the number of users are both large, where coded caching can provide the largest gain over the uncoded counterpart. However, within the problem setting  of\cite{MaddahAliNiesen:14} , the number of users $K$ accounted for in the placement phase is also the total number of simultaneous content requests in the delivery phase, and in some applications the number of simultaneous data requests can be quite small. In such scenarios, better understanding of the fundamental limits of the caching problem, when either $N$ or $K$ is small, in fact becomes rather important. 

The outer bound provided in \cite{MaddahAliNiesen:14} was obtained through a cut-set argument, and it is generally suspected to be not tight\footnote{The author wishes to acknowledge the conservation with Dr. Urs Niesen regarding this point.}. However, stronger outer bounds appear difficult to find analytically. In this short note, we provide an improved outer bound for the case with $N=3$ files and $K=3$ users, which is obtained using the computational approach developed in \cite{Tian:JSAC13}. This outer bound is included in the online collection of ``Solutions of Computed Information Theoretic Limits (SCITL)''  at \cite{TianWebpage}, where the data files for the proof can be downloaded for further processing.

\section{An Improved Outer Bound for the $N=K=3$ Caching Problem}
\label{sec:main}

We use the problem definition and notation given in \cite{MaddahAliNiesen:14}: $M$ is the capacity of the local memory cache size, and $R$ is the multicast rate in the delivery transmission; each file is assumed to have unit size, {\em i.e.,} $F=1$, since normalizing by $F$ does not cause any essential loss here. The main result of this note is the following theorem.
 
\begin{theorem}
The memory-delivery-rate tradeoff for the $N=K=3$ coded caching problem must satisfy: 
\begin{align}
M\geq 0, \quad 3M+R\geq 3, \quad 6M+3R\geq 8,\quad M+R\geq2,\nonumber\\
12M+18R\geq 29,\quad 3M+6R\beta\geq 8,\quad M+3R\geq 3, \quad R\geq 0. \label{eqn:bounds}
\end{align}
\end{theorem}

\begin{figure}
\centering
\includegraphics[width=10cm]{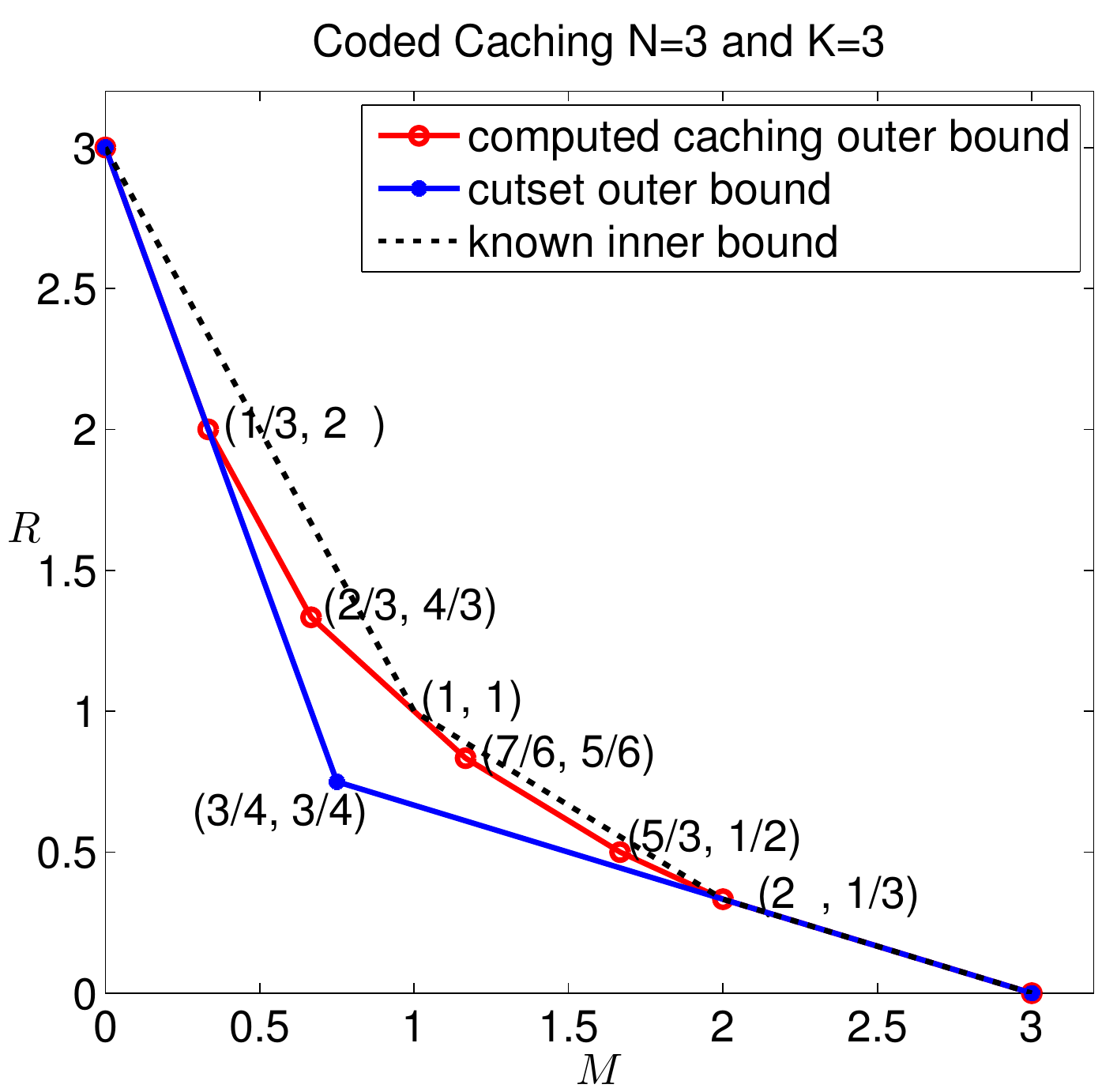}
\caption{The memory-transmission tradeoff for $N=K=3$. \label{fig:region}}
\end{figure}

This region is illustrated in Fig. \ref{fig:region}. We note that the point $(1,1)$, which was shown to be achievable in \cite{MaddahAliNiesen:14}, is optimal since it is on the boundary of the outer bound. 
The first two and the last two inequalities in (\ref{eqn:bounds}) are already known in \cite{MaddahAliNiesen:14}. The remaining ones are new, and we discuss them in some details in the next section. For reference, the inner bound given in \cite{MaddahAliNiesen:14} through a centralized placement algorithm is also plotted.  We shall not provide the full details of  the proof in this note, since some of them can be rather large (in a table form, or long when written down as chains of inequalities). However, one of the inequalities is discussed in some depth in Section \ref{sec:translation} to illustrate the manner this bound is obtained, and interested readers are referred to \cite{TianWebpage} for more details.

\section{Symmetry Structure}

In this section we discuss the symmetry we utilized to reduce the complexity of the computation, which is important to understand the tabulation based proof. In the sequel we use $W_i$ to denote the $i$-th (random) file, $Z_i$ to denote the content stored at the $i$-th user. The multicasted message in the delivery phase is written as $X_{i,j,k}$, meaning that it is the message when the first user demands the $i$-th file, the second user demands the $j$-th file, and the third user demands the $k$-th file. The files are mutually independent.

Similar as in \cite{Tian:JSAC13}, it can be shown that without loss of generality we can consider only symmetric codes. However, here the symmetry structure is different from that in \cite{Tian:JSAC13}. In the computational approach, we utilize the following type of symmetry to reduce the computation. Let three sets of random variables be given as 
\begin{align}
\mathcal{W}\subseteq \{W_0,W_1,W_2\},\quad \mathcal{Z}\subseteq \{Z_0,Z_1,Z_2\},\quad \mathcal{X}=\{X_{s_0,s_1,s_2}:s_0,s_1,s_2\in \{0,1,2\}\}.
\end{align}
Let a permutation function be defined as $\pi(\cdot)$ on the set of $\{0,1,2\}$. The permutation operates on the set of $\mathcal{Z}$ as follows
\begin{align}
\pi_z(\mathcal{Z})\triangleq\{Z_{\pi(i)}:Z_i\in \mathcal{Z}\}
\end{align}
and on the set $\mathcal{X}$ as follows
\begin{align}
\pi_x(\mathcal{X})\triangleq\{X_{s_0,s_1,s_2} :X_{s_{\pi(0)}, s_{\pi(1)},s_{\pi(2)}}\in\mathcal{X}\}
\end{align}
For example the permutation function $\pi(0)=1,\pi(1)=2,\pi(2)=0$ maps $\{Z_1\}$ to $\pi_z(\{Z_1\})=\{Z_2\}$, but maps any $X_{s_1,s_2,s_0}$ to $X_{s_0,s_1,s_2}$, and thus the set $\{X_{0,1,2},X_{2,1,0}\}$ to $\{X_{2,0,1},X_{0,2,1}\}$. 

We call a given caching code symmetric, if for any $\mathcal{W},\mathcal{Z},\mathcal{X}$, and any permutation $\pi$, we have 
\begin{align}
H(\mathcal{W},\mathcal{Z},\mathcal{X})=H(\mathcal{W},\pi_z(\mathcal{Z}),\pi_x(\mathcal{X})).
\end{align} 
As a reality check, consider $H(W_1,Z_1,X_{0,1,2})$ under the aforementioned permutation: it should be equal to $H(W_1,Z_2,X_{2,0,1})$; notice that $W_1$ is a function of $(Z_1,X_{0,1,2})$ and under the permutation, $W_1$ is still a function of $(Z_2,X_{2,0,1})$. Intuitively, we prove that symmetric codes are without loss of optimality based on the idea of time (space) sharing: encode $1/6$ of the three files and place the coded cache in a permuted order at the three users, which will remove any asymmetry in the code\footnote{This only captures partially the symmetry in the problem, however it is sufficient to establish the outer bound.}. We plan to explore the symmetry structure further in a subsequent work.

\section{From Tabulation to Chains of Inequalities}
\label{sec:translation}

\begin{table}[ct]
\begin{center}
\caption{The entropy terms used in the proof.}
\label{tab:correspondence}
\begin{tabular}{|c|c|}
\hline
$T_{ 1}$ & $F=1$ \\
$T_{ 2}$ & $H(X_{2,1,0})$ \\
$T_{ 3}$ & $H(W_{2},X_{2,1,0})$ \\
$T_{ 4}$ & $H(W_{2},X_{2,0,1},X_{2,1,0})$ \\
$T_{ 5}$ & $H(Z_{2})$ \\
$T_{ 6}$ & $H(Z_{0},X_{2,1,0})$ \\
$T_{ 7}$ & $H(Z_{0},X_{2,0,1},X_{2,1,0})$ \\
$T_{ 8}$ & $H(W_{2},Z_{0})$ \\
$T_{ 9}$ & $H(W_{2},Z_{1},X_{2,1,0})$ \\
\hline
\end{tabular}
\end{center}
\end{table}

\begin{table*}[tcb]
\setlength{\tabcolsep}{4pt}
\begin{center}
\caption{Proof of the inequality $M+R\geq 2$ with terms defined in Table \ref{tab:correspondence}.}
\label{table:cancellation}
\begin{tabular}{|ccccc cccc|}
\hline
$T_1$  &$T_{ 2}$  &$T_{ 3}$  &$T_{ 4}$  &$T_{ 5}$  &$T_{ 6}$  &$T_{ 7}$  &$T_{ 8}$  &$T_{ 9}$  \\\hline
            &          &          &          &          &$  2$     &$ -1$     &$ -1$     &           \\
$ -3$     &          &$ -1$     &$  1$     &          &          &          &          &$1$     \\
$ -1$     &          &$1$     &          &          &          &          &$1$     &$ -1$     \\
          &$2$     &          &          &$  2$     &$ -2$     &          &          &           \\
          &          &          &$ -1$     &          &          &$1$     &          &           \\
\hline     
\hline
$-4$ &$2$  &        &        &    $2$    &         &       &        &       \\\hline
\end{tabular}
\end{center}
\end{table*}

To illustrate the proof obtained using the computational approach, consider the inequality $M+R\geq2$, which is particularly simple to prove\footnote{In fact, the author was informed by both Dr. Tie Liu and Dr. Vaneet Aggarwal that they had independently obtained this particular inequality analytically.}. 
This proof is a direct translation of the solution obtained using the computational approach, which are given in Table \ref{tab:correspondence} and Table \ref{table:cancellation}, where we use $F=1$ to denote the unit of the file size. 

We can alternatively write the following chain of inequalities
\begin{align}
 2H(Z_0)+2H(X_{2,1,0})&\geq 2H(Z_0,X_{2,1,0})\nonumber\\
&\stackrel{(s)}{=} H(Z_0,X_{2,1,0})+H(Z_0,X_{2,0,1})\nonumber\\
&\stackrel{(a)}{=}H(Z_0,W_2,X_{2,1,0})+H(Z_0,W_2,X_{2,0,1})\nonumber\\
&=2H(Z_0,W_2)+H(X_{2,1,0}|Z_0,W_2)+H(X_{2,0,1}|Z_0,W_2)\nonumber\\
&\geq 2H(Z_0,W_2)+H(X_{2,1,0},X_{2,0,1}|Z_0,W_2)\nonumber\\
&=2H(W_2)+2H(Z_0|W_2)+H(X_{2,1,0},X_{2,0,1}|Z_0,W_2)\nonumber\\
&\stackrel{(b)}{=} 2+H(Z_0|W_2)+H(Z_0,X_{2,1,0},X_{2,0,1}|W_2)\nonumber\\
&\geq 2+H(Z_0|W_2)+H(X_{2,1,0},X_{2,0,1}|W_2)\nonumber\\
&\stackrel{(s)}{=} 2+H(Z_1|W_2)+H(X_{2,1,0},X_{2,0,1}|W_2)\nonumber\\
&\geq 2+H(Z_1,X_{2,1,0},X_{2,0,1}|W_2)\nonumber\\
&\stackrel{(c)}{\geq} 2+H(W_0,W_1|W_2)=4, \label{eqn:chain}
\end{align}
where (s) is for reason of symmetry, and (a) is because the message $X_{2,1,0}$ together with the coded content cached at user-$0$, {\em i.e., $Z_0$}, can recover $W_2$, (b) is because of the assumption that each file has unit size as well as the chain rule, and (c) is because the coded content cached at user-$1$, {\em i.e., $Z_1$}, together with $X_{2,1,0}$ can recover $W_1$, and $Z_1$ together with $X_{2,0,1}$ can also recover $W_0$. Note the five inequalities in (\ref{eqn:chain}) correspond to the five rows in Table \ref{table:cancellation}, though not in the same order, and not the exact same form because of the symmetry structure and application of certain chain rule simplifications.

\section{Conclusion}

An improved outer bound is given for the caching problem when $N=K=3$ in this note. This result is part of the online collection of ``Solutions of Computed Information Theoretic Limits (SCITL)'' hosted at \cite{TianWebpage}, which hopefully in the future can serve as a data depot for information theoretic limits obtained through computational approaches. Some results in this collection requires non-trivial variation of the approach outlined in \cite{Tian:JSAC13}, the details of which will be presented elsewhere. 

\bibliographystyle{IEEEbib}

\end{document}